# Direct observation of nanometer-scale orbital angular momentum accumulation

Juan Carlos Idrobo[1,2], Ján Rusz[3], Gopal Datt[3], Daegeun Jo[3,4], Sanaz Alikhah[3], David Muradas[3], Ulrich Noumbe[3], M. Venkata Kamalakar[3], and Peter M. Oppeneer[3,4]

[1]Materials Science & Engineering, University of Washington, 302 Roberts Hall, Seattle, WA 98195-2120, USA
[2]Physical and Computational Sciences Directorate, Pacific Northwest National Laboratory, Richland, WA, 99354, USA
[3]Department of Physics and Astronomy, Uppsala University, P.O. Box 516, SE-75120 Uppsala, Sweden
[4]Wallenberg Initiative Materials Science for Sustainability, Uppsala University, SE-75120 Uppsala, Sweden

**Abstract:**
Conversion of charge to orbital angular momentum through the orbital Hall effect (OHE) holds transformative potential for the development of orbital-based electronics, however, it is challenging to directly observe the electrically generated orbital accumulation. Here, we detect the OHE by directly quantifying the orbital accumulation along the edges of a titanium thin film using a scanning transmission electron microscope. We measure the Ti L-edge using electron energy-loss spectroscopy with nanometer resolution and find a sizable orbital accumulation at the sample's outer perimeters, consistent with all signatures expected for the OHE, and determine an orbital diffusion length $\ell_o \approx 7.3$ nm. Our data points to a surprising dependence of the orbital diffusion length on the nano-structural morphology.

# 1. Introduction

The spin Hall effect (SHE) converts a charge current into a transverse spin current [1]. The SHE has emerged as a key enabling ingredient for spin-based electronic elements, because the SHE can induce a spin-orbit torque on the magnetization in heavy-metal/ferromagnetic metal bilayers [2] that is strong enough to reverse the magnetization of the ferromagnetic layer [3,4]. As the SHE is proportional to the relativistic spin-orbit coupling (SOC), it is typically large for heavy 5d metals, such as W or Pt [2]. Apart from the spin angular momentum, electrons also carry orbital angular momentum. It was predicted theoretically that there should exist also an orbital Hall effect (OHE) [5]. *Ab initio* calculations predicted that, since the OHE does not require SOC [6-9], the OHE would be considerably larger than the SHE for light 3d-transition metals [10,11]. These predictions offer exciting prospects for a possible transformation from spin-based electronics, using heavy metals, to 'orbitronics' by employing the electrons' orbital momentum instead of their spin momentum, in light elements [12,13]. However, despite the intriguing theoretical predictions, a convincing proof of the OHE's existence has been difficult to achieve.

Recently, two techniques, the magneto-optical Kerr effect (MOKE) and Hanle magnetoresistance, were employed to detect the OHE in thin films of the light 3d-transition metals Ti, Cr, and Mn [14-16]. Although these measurements provide evidence for the existence of the OHE, there remain many particulars of the OHE that are not yet understood. Specifically, the recent measurements do not give information on how the current-induced accumulation of spin and orbital angular momenta occurs on the nanometer scale. MOKE probes a sample area of a few $\mu m^2$, yet the transport and accumulation of spin and orbital momenta is expected to occur on a much shorter, nanometer length scale in metals. A second intensively debated topic is the size of the orbital diffusion length $\ell_o$. Since spin and orbital responses always occur together in the presence of SOC, it is nontrivial to distinguish them and to uniquely extract the value of $\ell_o$ from experiments. Widely different values for orbital diffusion lengths were reported recently, varying from below 2 nm for Mn [16], 2.9 nm for Cu [17], 6.6 nm for Cr [15], 50-74 nm for fcc Ti [14,18], and 80 nm for the heavy-metal W [19]. Some of these orbital diffusion lengths appear antagonistic to spin diffusion lengths, for example, those that are known to be long in Cu but rather short in W [20]. To resolve these issues, a direct observation of the OHE and its orbital diffusion length is desirable, which would require probing techniques that have both nanometer-scale sensitivity and can unambiguously distinguish between spin and orbital angular momenta.

# 2. Nanometer-scale detection of orbital angular momentum

We detect here the orbital momentum due to the OHE using a scanning transmission microscope (STEM). The key to observing the OHE with nanometer resolution lies in the electron-optical alignment of the STEM, ensuring that the electron momentum transfer mimics the role of polarized light in analogous x-ray magnetic circular dichroism experiments [21]. Using this principle, magnetism can be measured with electron energy-loss spectroscopy (EELS) in electron magnetic circular dichroism (EMCD) experiments [22-24]. Advantages of EMCD in STEM are the near atomic spatial resolution and its element specificity which is achieved when specific core levels are excited.

We employed a nanometer-sized electron beam to investigate spin and orbital angular momentum accumulation in a single light 3d-element, hcp Ti. Figure 1a shows schematically our measurement principle. In our experiments, performed at room temperature, we used an electron beam with an equivalent of 2-nm spatial resolution that was transmitted through the 24-nm Ti thin film on silicon nitride. We positioned the EELS collection aperture off-axis at a distance sufficient to avoid interference from the direct electron beam. We placed the electron probe at one edge of the sample, applied a bias voltage, and acquired a spectrum. Then, we reversed the bias direction and obtained another spectrum. The EMCD is obtained as the difference between these two spectra. This process was replicated for the opposite edge. The resulting EMCD signal is then analogous to that obtained from two spectra acquired for opposite magnetizations, but without the need to flip the sample or switch the magnetic field of the objective lenses.

Through the applied bias voltage, the SHE and OHE induce a transversal angular momentum flow that leads to an accumulation of both spin and orbital angular momenta on the edges of the sample, but with opposite sign for the top and bottom edge of the sample, see Fig. 1b. The key idea behind the STEM experiment is to scan with nanometer resolution the regions near the perimeters of the sample, where a large spin or orbital accumulation is expected. In these regions, we acquired EELS at the Ti $L_{3,2}$-edge, which corresponds to 2p to 3d excitations and thus dominantly probes the magnetic response of the Ti 3d electrons.

Figure 1c shows an optical microscope image of the 4.5 µm-wide Ti film on an insulating silicon-nitride membrane of a STEM chip, with electrodes fabricated by electron-beam lithography. The Ti film was coated with 5 nm alumina to prevent oxidation (see Supplementary Materials for details). We focus our measurement on the two perimeters of the film, the top and bottom edge, see Fig. 1b, on which spin and orbital momentum accumulation with opposite signs due to the SHE and OHE, respectively, is expected.

The presence of spin or orbital polarization on the Ti atom leads to a particular modification of the L-edge fine structure. Figures 2a and 2b show schematically the EELS spectra for opposite bias voltages and their intensity difference, i.e., the EMCD, with respect to that of an unpolarized atom, as expected for a pure spin polarization (Fig. 2a), and a pure orbital polarization (Fig. 2b). An EMCD signal exhibiting pure spin polarization manifests as two equally large peaks at the $L_3$ and $L_2$ edges, with opposite signs. Conversely, a pure EMCD orbital polarization signal is characterized by two peaks at the $L_3$ and $L_2$ edges, both having the same sign (Fig. 2b) [25].

Figures 2c and 2d show representative EELS and EMCD spectra measured at the top and bottom perimeters, respectively. To obtain statistically relevant data, we averaged 120 spectra obtained along the perimeter from the acquired spectral images. We find that the voltage difference leads to a typical modification of the EELS spectrum: an increase or decrease of the whole spectrum at the $L_{3,2}$ edges. The resulting EMCD spectra are shown by the red and blue shaded areas in Figs. 2c and 2d, respectively. As can be recognized from Figs. 2c and 2d, the sign of the EMCD spectrum is reversed for the top and bottom edges. This presents the first evidence that an applied voltage leads to an antisymmetric modification of the EMCD spectrum for the two opposite sides of the sample.

Considering next the shape of the EMCD, we note that the measured shape corresponds closely to what is expected for an induced orbital polarization on the Ti atom (Fig. 2b), providing thus the second direct evidence for orbital accumulation. The presence of a small induced spin moment cannot be straightforwardly observed from the EMCD spectrum, but, in view of the shape of the spectra, we expect it to be negligible.

To achieve a quantitative analysis of the EMCD spectra as a function of bias voltage $V$, we extract a quantity that is proportional to the local orbital momentum on Ti by using the orbital momentum sum rule for EELS [25,26],

$$\int_{L_{3,2}} \left[I_{L_{3,2}}^{V+}(E) - I_{L_{3,2}}^{V-}(E)\right]dE = \int \Delta I_{L_{3,2}}(E) dE = \frac{N}{2n_h} K \langle \hat{L}_z \rangle, \quad (1)$$

where $I_{L_{3,2}}^{V\pm}(E)$ is the bias-voltage dependent EELS intensity, $\langle \hat{L}_z \rangle$ is the expectation value of the orbital angular momentum, $n_h$ is the number of holes in the 3d shell, and $N$ is the integral of the total EELS spectrum at the L-edge, i.e., $N = \int \frac{1}{2}\left(I_{L_{3,2}}^{V+}(E) + I_{L_{3,2}}^{V-}(E)\right)dE$. The coefficient $K$ depends on the experimental geometry [26,27]. Its value is unknown here, but constant, and thus the spectral integral (1) gives a direct proportionality to $\langle \hat{L}_z \rangle$. The local orbital magnetic moment is given by $m_L = -(\mu_B/\hbar)\langle \hat{L}_z \rangle$.

The OHE depends linearly on the applied field, and the induced angular momentum is therefore expected to increase linearly with the applied voltage. To verify this preposition, we selected two nanometer-sized areas on opposite sides of the sample, see Fig. 1c, and measured the EMCD spectra for various voltages. We then apply the orbital momentum sum rule to extract the local orbital momentum. In Fig. 2e we show the normalized EMCD integral (i.e., $\int \Delta I_{L_{3,2}}(E) dE$ divided by its maximum measured value) versus the bias voltage. This quantity is expected to be proportional to the orbital momentum. The encircled red and blue data points correspond to the EMCD spectra presented in Figs. 2c and 2d. We find that the induced orbital momentum scales *linearly* with voltage for both the top and bottom edges of the sample. We performed least-square fits with a linear function to the data for each perimeter independently and as an aggregated dataset. We find that in all three cases the slopes are the same and with intercepts near zero very close to $V = 0$. Note that the two edges represent in fact two physically independent measurements. It is therefore remarkable that, within error bars, the same slope is obtained. This strongly supports that the opposite orbital momenta measured at the opposite sample borders are generated by the same charge-to-orbital conversion process, which provides our third direct evidence for the direct detection of orbital accumulation.

### 3. Orbital angular momentum accumulation and decay

Figures 3a and 3b depict Ti L-edge EELS signals, integrated from 451 eV to 470 eV, illustrating the nanometer structure of the top and bottom perimeters, where the edge locations are defined as the positions where the second derivative of the integrated EELS signal (in *y*-direction) vanishes. Interestingly, a structural difference between the two edges is observed, with the top edge being relatively straight and uniform. The bottom edge is, however, much rougher, with spatially localized variations in both the sample's thickness and irregularities along the perimeter. Such structural differences could originate from the ~10 nm electron spot size employed in the e-beam lithography patterning, as well as from the grain size in the fabricated titanium stripe (see Supplementary Materials).

To measure the decay of orbital accumulation as a function of distance from the edges, we averaged 60 EMCD spectra along the edges in the *x*-direction (see Fig. 1c). For each distance from the edge, we applied a rolling average of two probe positions to enhance the signal-to-noise ratio, varied the bias voltage, and then employed the orbital momentum sum rule to obtain its magnitude, proportional to the orbital momentum. Figure 3c shows the extracted orbital signals (square symbols) as a function of this distance. As expected, the accumulated orbital polarization decays with distance from the edge.

To extract the orbital decay length, we note that the measured quantity is a convoluted signal of the electron beam (modeled here as a Gaussian function), the shape of the edge, and the exponentially decaying orbital accumulation $I_L^0 e^{-y/\ell_o}$ (see Supplementary Materials). The orbital diffusion length $\ell_o$ is defined here exactly in the same way as the spin diffusion length in the Valet-Fert model [27,28]. The anticipated orbital accumulation decay is given by

$$I_L(y) = I_L^0 e^{\sigma^2/2\ell_o^2} e^{-y/\ell_o} \frac{d}{2}\left(1 + \text{erf}\left[\frac{y - \sigma^2/\ell_o}{\sqrt{2\sigma^2 + w^2}}\right]\right), \qquad (2)$$

where $\sigma$ defines the width of the 2-nm electron beam ($\sigma \simeq 0.85$ nm), $d$ is the sample thickness and $w$ describes the width of the edge, with a total edge width of $\simeq 4w$. The value of $w$ is about 4 – 5 nm, but note that this value can vary locally due to nanotexture of the edge. Since $w$ could be comparable to $\ell_o$ it is explicitly taken into account in the model. Using Eq. (2) we fit the measured data, shown by the solid lines in Fig. 3c. The intensity $I_L^0$ represents merely the height of the curve, its spatial dependence is governed by $\ell_o$. We obtain that the top-edge data can be described well with the response function (2), giving $\ell_o \approx 7.3$ nm, with an error of 0.5 nm estimated from the fit. For the bottom edge, the data exhibit greater scatter and larger deviations but can still be reasonably well approximated by the fit function. We obtain an $\ell_o \approx 32$ nm for the bottom edge, but with an estimated error of ~17 nm. The noticeable difference between these values provides the first indication that the orbital diffusion length depends sensitively on the detailed nanostructure of the thin metallic film.

## 4. Discussion and conclusions

The detected EMCD signal fulfills the expected characteristics of orbital accumulation due to the OHE. Namely, the magnetic signal has opposite sign on the top and bottom edges (Figs. 2c,2d) and increases linearly with identical slope when the applied bias voltage is increased (Fig. 2e). Moreover, the recorded EMCD, shown in Figs. 2c,d unambiguously establishes that orbital momentum accumulation is detected here. We can, however, not exclude that a small spin accumulation is present as well, but we do not find a detectable spectral shape as expected for spin accumulation (see Fig. 2a). This finding is consistent with our *ab initio* density-functional theory calculations [29] that predict a minute spin Hall conductivity for hcp Ti, $\sigma_{\text{SHE}} \simeq -10 \left(\frac{\hbar}{e}\right)[\Omega \text{ cm}]^{-1}$, in contrast to a three-orders of magnitude larger orbital Hall conductivity of $\sigma_{\text{OHE}} \simeq 4 \times 10^3 \left(\frac{\hbar}{e}\right)[\Omega \text{ cm}]^{-1}$, while both quantities depend only moderately on the crystal orientation (see [11] and Supplementary Materials). In addition, the *ab initio* calculated normal Hall effect of Ti ($\sigma_{\text{NHE}} \simeq 8.8 \times 10^{-2} [\Omega \text{ cm T}]^{-1}$ is also several orders of magnitude smaller than the normal conductivity $\sigma_{\text{DC}} \simeq 7.2 \times 10^3 [\Omega \text{ cm}]^{-1}$) and therefore, it is not expected to play a role in our measurements.

The orbital diffusion length is the key quantity for the transport of orbital information in devices. This quantity has been difficult to determine directly. Very disparate values for $\ell_o$ were reported recently, varying from below 2 nm for Mn [16], to 50-74 nm for fcc Ti [14] and 80 nm for the heavy-metal W [19]. Measurements of current-induced torques in ferromagnetic/nonmagnetic metallic heterostructures have provided evidence of the existence of orbital currents [18,30-32], but it is nontrivial to deduce whether the resulting torque on the free magnetic layer is due to a generated spin or orbital polarization. Moreover, it is complicated to pinpoint from which layer or interface in the heterostructure the spin or orbital momentum originates. Due to the presence of multiple interfaces, various effects will occur, such as spin [33] and orbital [34] Rashba-Edelstein effects, in addition to self-torques in the magnetic layer [35]. All these will contribute simultaneously to the torque on the magnetization, which makes it nontrivial to attribute the resulting torque to uniquely one of the effects.

An alternative approach is to study terahertz emission caused by ultrafast demagnetization of a magnetic layer that delivers a femtosecond spin and orbital current pulse into an adjacent nonmagnetic layer. Recent THz emission measurements on CoPt/Cu bilayers revealed an orbital diffusion length of $\ell_o \approx 2.9$ nm for Cu [17], significantly smaller than its spin diffusion length [20]. Conversely, from THz measurements on Ni/W bilayers, an $\ell_o \approx 80$ nm was deduced for the heavy-metal W, much longer than its spin diffusion length [19]. Further investigations to clarify how the THz signal relates to the spin and orbital diffusion lengths are evidently required, as discussed recently [36].

Hanle magnetoresistance and MOKE measurements on nanometer thin films of light 3d metals, such as Mn [16], fcc Ti [14], and bcc Cr [15], provide a straightforward way to detect current induced orbital accumulation. To interpret the MOKE measurements, one has, however, to resort to *ab initio* calculations of orbital magneto-optics to extract $\ell_o$, which may introduce uncertainties. In both the Hanle and MOKE measurements, the orbital diffusion length is deduced by measuring the responses for multiple films with various thicknesses. In contrast, here we observe directly the nanometer-scale decay of orbital polarization on a single Ti film.

The orbital diffusion lengths for the two sides of the Ti strip are remarkably different. The $\ell_o$ of 7.3 nm of the top edge is similar to that of Cr, $\ell_o \approx 6.6$ nm [15,37], but larger than those of both Cu [17] and Mn [16]. Recent measurements on Ni/Ti bilayers provided disparate values for titanium's orbital diffusion length, varying from $\ell_o \approx 5$ nm obtained from magnetoresistance measurements [38], to $50 \pm 15$ nm obtained from torque measurements [14,18], and to $\ell_o \approx 74 \pm 24$ nm obtained from MOKE measurements on pure Ti [14]. Importantly, we can for the first time directly correlate the orbital decay length with the local nanostructure of the sample, see Figs. 3a,b, where the longer decay length corresponds to a more ragged nanostructure. An explanation of the longer $\ell_o$ could plausibly be that the crystal field that quenches orbital momentum [14] is reduced due to a relatively rough nanostructure. Another possible scenario is that, at nanometer-scale interfaces between crystallites, the orbital Rashba-Edelstein effect [34] locally generates orbital momentum and thus provides an additional source of orbital polarization at distances away from the perimeter. It can furthermore be suspected that the current density at such places is not uniform, which might increase the influence of this additional orbital momentum source.

The orbital diffusion length measured at the top edge is relatively short, $\ell_o \approx 7.3$ nm, but it is likely that this value, too, depends on the local nanostructure, and therefore, the intrinsic orbital diffusion length of high-quality single-crystalline films could be even shorter. Our work charts thus a direction for future EMCD STEM measurements of meticulously prepared samples to shed further light on how diffusion lengths and sample quality are connected on the nanoscale. More broadly, the here-developed methodology bridges the gap between nanometer-scale structure and electrically induced angular momenta. Using atomic-sized electron probes in STEM opens up an exciting era where we can reveal, with elemental sensitivity, the interconnections between structural morphology and induced spin and orbital momenta at nanometer scales that are key elements for spintronic and orbitronic device operations.

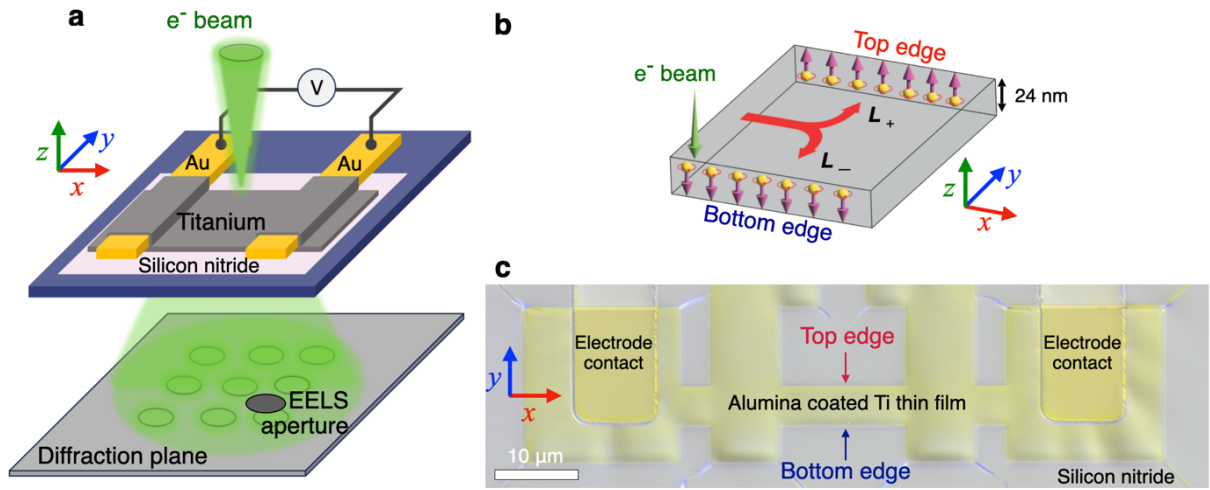

**Figure 1.** Measurement principle for detection of spin and orbital Hall effects in titanium. (a) An electron beam in a scanning transmission electron microscope (STEM) passes through the 24-nm Ti sample on a silicon-nitride membrane. EELS spectra are acquired off-axis for different bias voltages applied across the metal film. (b) Sketch of the expected accumulation of spin and orbital angular momenta due to the SHE and OHE, respectively, having opposite signs for the top and bottom edges of the sample. The 2-nm sized electron beam is used to scan with nanometer resolution the near-edge regions where a large spin or orbital accumulation is expected. (c) Optical microscope image of the alumina-coated Ti thin-film device. A two-dimensional set of spectrum images was acquired across and along the top and bottom edges of the 4.5-μm-wide Ti film for various bias voltages supplied by the electrodes.

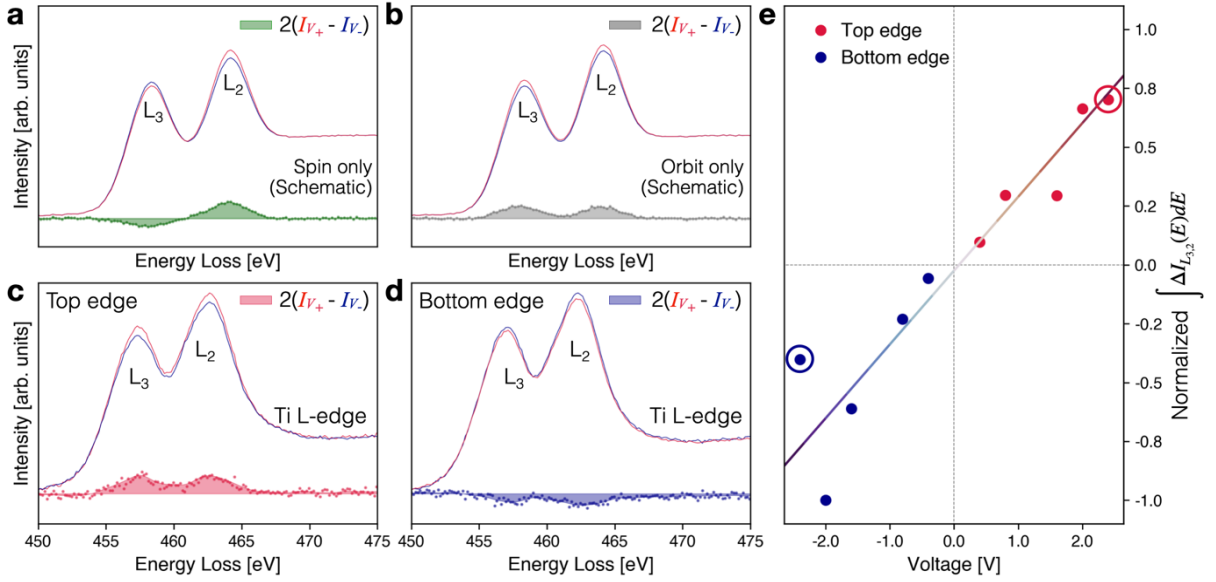

**Figure 2.** Orbital angular momentum identified with electron-energy loss spectroscopy. (a) Schematics of the shape of the EELS signals at the L-edge of Ti induced for opposite bias voltages and due to spin accumulation only. The corresponding EMCD spectrum is shown by the green shaded area. (b) Schematic shape of EELS signals and resulting EMCD (grey shaded) due to orbital accumulation only. (c) and (d) EELS spectra measured at the top and bottom edge, respectively, for bias voltages of +1.2 V and -1.2 V. The EMCD spectra acquired at both the top and bottom edges demonstrate a characteristic fingerprint of orbital accumulation (shown in Fig. 2b). (e) EMCD spectrum integrated over the whole L-edge, $\int \Delta I_{L_{3,2}}(E)dE$, as function of the bias voltage, and normalized to the value at maximum voltage. The integrated EMCD signal which is proportional to $\langle \hat{L}_z \rangle$ demonstrates linear dependence on applied voltage. The red and blue circles in (d) depict the data points corresponding to the spectra displayed in panels (b) and (c).

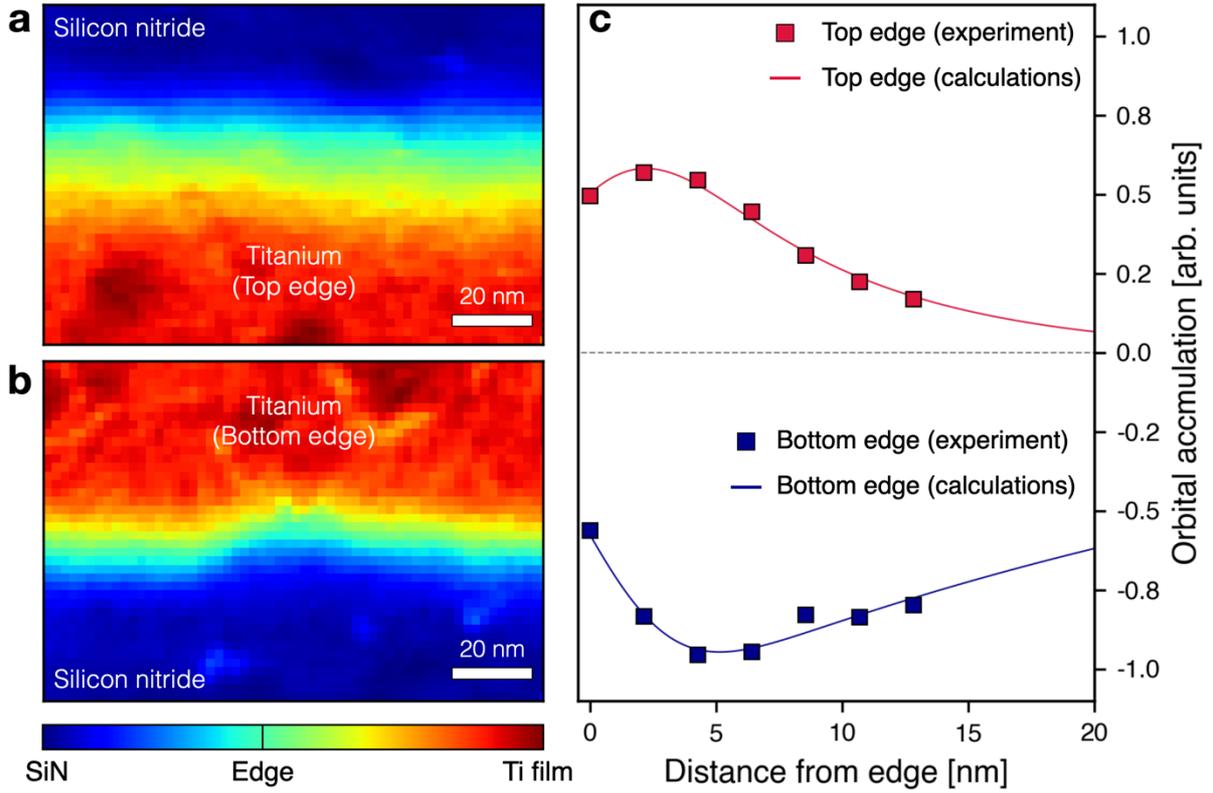

**Figure 3.** Nanostructure of the titanium film and measured orbital accumulation. (a) Integrated Ti L-edge EELS spectra obtained from regions of the top edge, and (b), of the bottom edge of the titanium sample (see Fig. 1c). The color coding represents the local thickness of the Ti film with a lateral resolution of about 2.1 by 2.1 nm$^2$. (c) The measured orbital momentum accumulation as a function of distance from the film's edge, for the top and bottom edge. Solid lines give a fit to the experimental data (squares) using the expected theoretical signal for orbital polarization decay, equation (2), to determine the orbital diffusion length, $\ell_o$.

**Acknowledgments**
We thank M. Berritta and D. Xiao for valuable discussions. This work was financially supported by the Swedish Research Council (VR, Grants No. 2021-03848 and 2021-05211), the Knut and Alice Wallenberg Foundation (Grant No. 2022.0079), the Olle Engkvist Foundation (Grants No. 200-0602 and 214-0331), and by the Wallenberg Initiative Materials Science for Sustainability (WISE) funded by the Knut and Allice Wallenberg Foundation. MVK acknowledges funding from the European Research Council (ERC), Project SPINNER. The EELS part of this research was supported by the Center for Nanophase Materials Sciences, which is a Department of Energy Office of Science User Facility. This research was conducted, in part, using instrumentation within ORNL's Materials Characterization Core provided by UT-Battelle, LLC under Contract No. DE-AC05-00OR22725 with the U.S. Department of Energy. This work was also partly funded under the Laboratory Directed Research and Development Program at Pacific Northwest National Laboratory, a multiprogram national laboratory operated by Battelle for the U.S. Department of Energy (JCI). The calculations were enabled by resources provided by the National Academic Infrastructure for Supercomputing in Sweden (NAISS) at NSC Linköping, partially funded by VR through Grant Agreement No. 2022-06725.

**Author contributions:** PMO initiated and guided the project. Experiments were conceived of and designed by JCI, JR, MVK and PMO. JR proposed to use EMCD for detection of OHE and designed the experimental geometry. GD, DM, UN, and MVK designed and prepared the samples. JCI performed the EMCD STEM measurements and the analysis of the EELS data. DJ performed the orbital accumulation fits and determined the orbital diffusion length. PMO performed modeling for the orbital diffusion length. SA and PMO performed the *ab initio* calculations. The paper was written by PMO with comments from other authors. All authors discussed the data and contributed to the manuscript.